\documentstyle[12pt,aasms4]{article}
\begin{document}

\vspace*{0.5cm}

\title{Relativistic Effects on Neutrino Pair Annihilation above a Kerr Black Hole
with the Accretion Disk}

\author{Katsuaki Asano$^1$ and Takeshi Fukuyama$^2$}

\vspace{2cm}

\affil{\altaffilmark{1}Department of Earth and Space Science,
Graduate School of Science,
Osaka University,
Toyonaka 560-0043, Japan}
\affil{\altaffilmark{2}Department of Physics, Ritsumeikan University,
Kusatsu, Shiga 525-8577, Japan}

\affil{\footnotesize e-mail: asano@vega.ess.sci.osaka-u.ac.jp, fukuyama@se.ritsumei.ac.jp}

\vspace{2cm}

\baselineskip=16pt

\abstract{Using idealized models of the accretion disk
we investigate
the relativistic effects on the energy deposition rate
via neutrino pair annihilation
near the rotation axis of a Kerr black hole.
Neutrinos are emitted from the accretion disk.
The bending of neutrino trajectories and the redshift due to
the disk rotation and gravitation are taken into consideration.
The Kerr parameter, $a$, affects not only neutrinos' behavior but also
the inner radius of the accretion disk.
When the deposition energy is mainly contributed by the neutrinos coming
from the central part, the redshift effect becomes dominant as $a$
becomes large and the energy deposition rate is reduced compared with
that neglecting the relativistic effects.
On the other hand, for small $a$
the bending effect gets dominant and makes energy increase
by factor 2 compared with that neglecting the relativistic effects.
For the disk with temperature gradient, the energy deposition rate
for a small inner radius of the accretion disk
is smaller than that estimated by neglecting the relativistic effects.
The relativistic effects, especially for large $a$,
play a negative role in avoiding the baryon contamination problem
in gamma-ray bursts.}

\vspace{0.5cm}

\noindent{\it Subject headings}: accretion, accretion disks---black hole physics---gamma rays: bursts

\newpage

\section{INTRODUCTION}

\indent

The light curves (Fruchter et al. 1999;
Harrison et al. 1999; Huang, Dai \& Lu 2000) and polarization
(Covino et al. 1999; Wijers et al. 1999) of gamma-ray burst afterglows
indicate anisotropic central engines.
These observations have been considered as evidence that gamma-ray bursts
are highly beamed.
One of the most probable candidates for the central engine of gamma-ray bursts
is the accretion disk around a black hole (Woosley 1993; Popham, Woosley \& Fryer 1999;
MacFadyen \& Woosley 1999; Ruffert \& Janka 1999).
This system may be formed
by the merging of two neutron stars,
the merging of a black hole and a neutron star,
or the failed supernovae (Woosley 1993).

In these systems the hot accretion disk emits neutrinos and antineutrinos.
The neutrino-antineutrino annihilation
into electrons and positrons
(hereafter neutrino pair annihilation)
is a possible energy source of gamma-ray bursts
(Paczy\'nski 1990; M\'esz\'aros \& Rees 1992; Jaroszy\'nski 1996;
Janka \& Ruffert 1996; Ruffert et al. 1997; Ruffert \& Janka 1998, 1999).
In order to get the gamma-ray bursts from the relativistic fireball
(Shemi \& Piran 1990; Rees \& M\'esz\'aros 1992;
M\'esz\'aros \& Rees 1993; Sari \& Piran 1995; Sari, Narayan \& Piran 1996),
the baryon density in the fireball must be extremely small.
Above the accretion disk,
the baryon density has the lowest value near the rotation axis.
The neutrino pair annihilation in this region
has the possibilities of making a clean fireball
and of solving the baryon contamination problem.
It is important to obtain the energy density profile
along the rotation axis in order to discuss the baryon contamination
problem and the beaming effect.

The energy deposition rate (hereafter EDR) due
to neutrino pair annihilation
(Goodman, Dar \& Nussinov 1987; Cooperstein, Van Den Horn \& Baron 1987;
Berezinsky \& Prilutsky 1987)
and the gravitational effects on it
(Jaroszy\'nski 1993; Ruffert \& Janka 1999; Salmonson \& Wilson 1999)
have been calculated.
In the recent study, Asano \& Fukuyama (2000; hereafter AF)
calculated EDR
by neutrinos emitted from the accretion disk.
They incorporated the gravitational effects due to the central Schwarzschild
black hole and concluded that the gravitational effects do not substantially change EDR.
However, AF did not consider the effects of the rotation
and temperature gradient of the accretion disk.
EDR is proportional to $T_{\rm eff}^9$,
where $T_{\rm eff}$ is the effective temperature of the accretion disk.
Therefore, in order to discuss EDR,
we have to include the temperature dependence on radius of the accretion disk.

The relativistic effects consist of three factors:
they are the gravitational redshift, the bending of neutrino
trajectories, and the redshift due to the disk rotation.
EDR is enhanced by the effect of neutrino bending.
However, the redshift due to the disk rotation and gravitation
reduces EDR.
Thus these effects complicate the estimate of EDR.
It is not obvious whether the bending effect becomes dominant
and EDR is increased or the redshift effect becomes dominant and EDR 
is decreased.
AF did not consider the Kerr black hole.
The rotation of the black hole affects the behavior
of neutrinos.
In addition the rotation of the black hole makes the inner radius of the accretion
disk smaller.
These effects may drastically change EDR
differently from the case for the Schwarzschild black hole.

In order to discuss quantitatively the central engine of
gamma-ray bursts, we need the comprehensive studies of the
formation, evolution, temperature dependence, opacity and geometry of
the accretion disk and with the mechanism of the energy deposition,
its efficiency, the environment around the fireball,
the evolution of the fireball and the efficiency of radiation in gamma-ray etc.,
most of which are affected by the general relativistic effects. 
We think that these problems will be solved finally by laborious
simulations.
However, no one reveals the reality of the central engine at the present stage
and can say anything affirmative on it.
In these situations we discuss qualitatively the problem limiting on the 
relativistic effects on neutrino pair annihilation which have not 
been treated seriously in numerical studies.
We do not deal with a specific model of gamma-ray bursts,
but derive general conclusions on EDR of gamma-ray bursts.
Our aim in this work is not to estimate quantitatively
the energy of gamma-ray bursts but to study semi-analytically
the relativistic effects on EDR
above the accretion disk around a rotating black hole.
For the above purpose we consider the most idealized situations and
compare the results under the same conditions except for 
the presence and absence of gravitation.
We calculate EDR along the rotation axis
where the baryon contamination is minimum.
The rotation and temperature gradient of the accretion disk
are taken into consideration.
We assume the most simple models of temperature gradient.

This paper is organized as follows. In section two we formulate
the algorithm of EDR calculation along the
rotation axis of a Kerr black hole.
The results caused by neutrinos emitted from the isothermal disk
and from the disk with $T_{\rm eff} \propto R^{-1}$ and $R^{-0.5}$
are given in section three.
The last section is devoted to conclusions.

\section{FORMULATION}

\indent

The formulation developed in this section is obtained by
modifying AF's work for the Schwarzschild black hole.
In the present case, we consider the following situations.
Around a rotating black hole, a hot thin accretion disk is formed
and emits a vast number of neutrinos.
Because we do not consider a specific model on the central engine of gamma-ray bursts,
we consider the most idealized situations irrelevant to
detailed models of the disk formation and so on.
The inner (outer) edge of this accretion disk is denoted by $R_{\rm in}$ ($R_{\rm out}$).
The geometrically thin disk is assumed to be sufficiently opaque
to neutrinos over the whole region.
We assume that the self-gravitational effects of the accretion disk is negligible
and neutrinos are gravitationally
affected only by the central black hole.
Of course, these idealizations may be far from the case of the realistic accretion disk.
However, we consider that this simple model is sufficient for qualitatively studying
the gravitational effects on EDR.
The metric around the rotating black hole is given by
\begin{eqnarray}
ds^2=g_{i j} dx^i dx^j=
\left( 1-\frac{r_{g} r}{\Sigma} \right) c^2 dt^2-\frac{\Sigma}{\Delta} dr^2 \nonumber \\
-\left( \frac{r_{g} r}{\Sigma} a^2 \sin^2{\theta}+r^2+a^2 \right) \sin^2{\theta} d\varphi^2
\nonumber \\
-\Sigma d \theta^2+2 \frac{r_{g} r}{\Sigma} a \sin^2{\theta} c dtd \varphi.
\end{eqnarray}
Here $\Sigma\equiv r^2+a^2 \cos^2{\theta}$, $\Delta\equiv r^2+a^2-r_{g} r$ and
$r_{g}=2GM/c^2$ is the Schwarzschild radius.
Also $a$ is the Kerr parameter and takes a value from $0$ to $r_{g}/2$.
In this field the eikonal for a massless particle 
is written as (Landau \& Lifshitz 1979)
\begin{equation}
\psi=-\omega_0 t+L \varphi+\psi_r(r)+\psi_{\theta}(\theta),
\label{eikonal}
\end{equation}
where $\omega_0$ and $L$ are constants and connected with
the energy and the angular momentum around the rotation axis of the black hole,
respectively.
$\psi_r(r)$ and $\psi_{\theta}(\theta)$ satisfy the equations
\begin{eqnarray}
\left( \frac{\partial \psi_r(r)}{\partial r} \right)^2=
\frac{1}{\Delta^2} \left[\frac{\omega_0}{c} (r^2+a^2)-a L \right]^2
-\frac{J^2}{\Delta}, \\
\left( \frac{\psi_{\theta}(\theta)}{\partial \theta} \right)^2=
J^2-\left[ \frac{a \omega_0}{c} \sin{\theta}-\frac{L}{\sin{\theta}} \right]^2,
\end{eqnarray}
where $J$ is a constant (the separation parameter)
and connected with the total angular momentum.
From equation (\ref{eikonal}),
we obtain the momentum of a neutrino by
$p_{i}=\hbar \partial \psi/\partial x^i$.

As was mentioned in Introduction, we are most interested in
EDR via neutrino pair annihilation
near the rotation axis, $\theta=0$.
The region of $\theta = 0$ allows analytical treatment because of its symmetric property.
In the absence of gravitation, AF indicated that
the $\theta$-dependence of EDR is weak for 
small $\theta$.
Although the $\theta$-dependence of the gravitational effects may not necessarily
be small, we calculate EDR at $\theta=0$
and take it as a rough standard in the vicinity of the rotation axis.

It should be remarked that $L=0$ is required for neutrinos to reach to $\theta=0$.
The eikonal equation only is not sufficient to obtain
$\varphi_{\nu}$, which is the angle from which meridian the particle
is appoaching the rotation axis.
Here and henceforth the subscript $\nu$ and $\bar{\nu}$ denote
the quantities related to neutrinos and antineutrinos, respectively.
Taking this $\varphi_{\nu}$ into consideration,
the inner product of the momenta, $p_{\nu}$ and $p_{\bar{\nu}}$, at $\theta=0$ becomes
\begin{equation}
(p_{\nu} \cdot p_{\bar{\nu}})=\frac{\varepsilon_{\nu} \varepsilon_{\bar{\nu}}}{c^2} 
\left( 1-\sin{\theta_{\nu}} \sin{\theta_{\bar{\nu}}}
\cos{(\varphi_{\nu}-\varphi_{\bar{\nu}})}-\cos{\theta_{\nu}} \cos{\theta_{\bar{\nu}}}
\right).
\end{equation}
Here the proper energy of neutrinos has been written as
\begin{equation}
\varepsilon_{\nu} = \hbar \omega_{0 \nu} \sqrt{\frac{r^2+a^2}{\Delta}}
= \varepsilon_{0 \nu} \sqrt{\frac{r^2+a^2}{\Delta}},
\end{equation}
where $\varepsilon_{0 \nu}$ is the energy observed at infinity.
The factor $\sqrt{(r^2+a^2)/\Delta}$ denotes the redshift.
$\theta_{\nu}$ is defined by
\begin{equation}
\sin{\theta_{\nu}} = \frac{\rho_{\nu} \sqrt{\Delta}}{r^2+a^2},
\label{sin}
\end{equation}
with
\begin{equation}
\rho_{\nu} \equiv \frac{c J_\nu}{\omega_{0 \nu}}.
\end{equation}

A neutrino is emitted from the disk at $(r,\theta)=(R,\pi/2)$,
and it arrives at a point $(r,0)$.
Then $\varphi_{\nu}$ can take values from
$0$ to $2 \pi$ at $(r,\theta)=(r,0)$ from the symmetric property.
On the other hand, a neutrino coming from $R_{\rm in}$ forms $\theta_{\rm m}$,
the minimum value of $\theta_{\nu}$, and that from
$R_{\rm out}$ forms $\theta_{\rm M}$, the maximum value of $\theta_{\nu}$.
These values are derived from equation (\ref{sin}) and
$\rho_{\nu}$ which is obtained by solving the trajectory equation
($\partial \psi/\partial J^2=$const.),
\begin{equation}
\int_0^{\pi/2} \frac{d \theta}{\sqrt{1-(a/\rho_{\nu})^2 \sin^2{\theta}}}
=\int_{\rm C} \frac{dr'}{r' \sqrt{(r'/\rho_{\nu})^2
\left(1+(a/r')^2 \right)^2-\left(1-r_g/r'+(a/r')^2 \right)}}.
\label{traj}
\end{equation}
Here, in the case in which a neutrino  passes through the nearest distance, $r_0$,
before it arrives at $\theta=0$,
the integration for $r'$ is performed from $r_0$ to $R$ and $r$.
When $r'$ varies monotonically,
the integration is performed from the smaller
to the larger of $r$ and $R$.
$r_0$ and $\rho_{\nu}$ are related by
\begin{equation}
\rho_{\nu}=\frac{r_0 (1+(a/r_0)^2)}{\sqrt{1-r_g/r_0+(a/r_0)^2}}.
\end{equation}

As was derived in AF, EDR
via neutrino pair annihilation for a distant 
observer is expressed as
\begin{eqnarray}
\frac{dE_0 (r)}{dt dV}&=&
\frac{21 \pi^4}{4} \zeta(5) \frac{K G_{\rm F}^2}{h^6 c^5} k^9 T_{\rm eff}^9(3 r_g) F(r),
\label{depo}
\end{eqnarray}
where
\begin{eqnarray}
F(r)&=&
\frac{1}{T_{\rm eff}^9(3 r_g) }
\left( \frac{r^2+a^2}{\Delta} \right)^4 \int_{\theta_{\rm m}}^{\theta_{\rm M}}
d \theta_{\nu} \sin{\theta_{\nu}}
\int_{\theta_{\rm m}}^{\theta_{\rm M}} d \theta_{\bar{\nu}} \sin{\theta_{\bar{\nu}}}
\int_0^{2 \pi} d \varphi_{\nu} \int_0^{2 \pi} d \varphi_{\bar{\nu}} \nonumber \\
&& \times T_{0}^5(R_{\nu}) T_{0}^4(R_{\bar{\nu}})
\left[ 1-\sin{\theta_{\nu}} \sin{\theta_{\bar{\nu}}}
\cos{(\varphi_{\nu}-\varphi_{\bar{\nu}})}-\cos{\theta_{\nu}} \cos{\theta_{\bar{\nu}}}
\right]^2.
\label{fr}
\end{eqnarray}
Here the following notations and remarks are in order. 
By the temperature at $R=3 r_g$, we have
represented the effective temperature of the disk $T_{\rm eff}$
which is observed in the comoving frame.
$T_0$ is the temperature of the disk observed at infinity.
$dt dV$ denotes $\sqrt{-g} d^4 x$.
The dimensionless parameter $K$ is 0.12 for $\nu_{\rm e}$ and 0.027 for
$\nu_{\rm \mu}$ and $\nu_{\rm \tau}$.
The Fermi constant $G_{\rm F}^2$ is $5.29 \times 10^{-44} {\rm cm^2\,
MeV^{-2}}$. $k$ and $h$ are  the Boltzmann  and the Planck
constants, respectively.
The factor of the redshift in equation (\ref{fr})
and the equation to obtain $\theta_{\rm m}$ and $\theta_{\rm M}$
are different from those for the Schwarzschild metric discussed in AF.
Furthermore, although it was assumed in AF that $T_0$ does not depend on $R$,
we have here assumed that $T_0$ depends on $R$.
From an angle $\theta_{\nu}$ at a point $(r,0)$,
we can trace back a trajectory of a neutrino
to the emitted position $R_\nu$ on the disk.
Thus $R_\nu$ and, therefore, $T_0(R_\nu)$ depend on $\theta_{\nu}$, and $T_0$ is
involved in the integrand of $\theta_{\nu}$.
This $\theta_{\nu}$-dependence of $T_0$ is obtained by the numerical
calculation of equation (\ref{traj}).

In the locally nonrotating frame (Bardeen, Press, \& Teukolsky
1972), a neutrino with $L=0$ moves normally to the direction of
the disk motion.
In that case, the temperature suffers the redshift due to the rotation 
of the disk and gravitation.
Namely,
\begin{equation}
T_0(R)=\frac{T_{\rm eff}(R)}{\gamma}
\sqrt{g_{tt}-\frac{g_{t \varphi}^2}{g_{\varphi \varphi}}}
=\frac{T_{\rm eff}(R)}{\gamma} \sqrt{\frac{R^2+a^2-r_g R}{R^2+a^2+(r_g/R) a^2}},
\end{equation}
where $\gamma \equiv 1/\sqrt{1-v^2/c^2}$ and 
\begin{equation}
\frac{v^2}{c^2}=\frac{r_g R (R^2+a^2-a \sqrt{2 r_g R})^2}
{2 (R^2+a \sqrt{r_g R/2})^2 (R^2+a^2-r_g R)},
\end{equation}
(Chandrasekhar 1983).

Integrating it over $\varphi_{\nu}$ and $\varphi_{\bar{\nu}}$, $F(r)$ becomes
\begin{eqnarray}
F(r)&=&
\frac{2 \pi^2}{T_{\rm eff}^9(3 r_g) }
\left( \frac{r^2+a^2}{\Delta} \right)^4
\left[ 2 \int_{\theta_{\rm m}}^{\theta_{\rm M}} d \theta_{\nu}
T_{0}^5(\theta_{\nu}) \sin{\theta_{\nu}}
\int_{\theta_{\rm m}}^{\theta_{\rm M}} d \theta_{\bar{\nu}}
T_{0}^4(\theta_{\bar{\nu}}) \sin{\theta_{\bar{\nu}}} \right. \nonumber \\
&&+\int_{\theta_{\rm m}}^{\theta_{\rm M}} d \theta_{\nu}
T_{0}^5(\theta_{\nu}) \sin^3{\theta_{\nu}}
\int_{\theta_{\rm m}}^{\theta_{\rm M}} d \theta_{\bar{\nu}}
T_{0}^4(\theta_{\bar{\nu}}) \sin^3{\theta_{\bar{\nu}}} \nonumber \\
&&+2 \int_{\theta_{\rm m}}^{\theta_{\rm M}} d \theta_{\nu}
T_{0}^5(\theta_{\nu}) \cos^2{\theta_{\nu}} \sin{\theta_{\nu}}
\int_{\theta_{\rm m}}^{\theta_{\rm M}} d \theta_{\bar{\nu}}
T_{0}^4(\theta_{\bar{\nu}}) \cos^2{\theta_{\bar{\nu}}} \sin{\theta_{\bar{\nu}}} \nonumber \\
&& \left. -4 \int_{\theta_{\rm m}}^{\theta_{\rm M}} d \theta_{\nu}
T_{0}^5(\theta_{\nu}) \cos{\theta_{\nu}} \sin{\theta_{\nu}}
\int_{\theta_{\rm m}}^{\theta_{\rm M}} d \theta_{\bar{\nu}}
T_{0}^4(\theta_{\bar{\nu}}) \cos{\theta_{\bar{\nu}}} \sin{\theta_{\bar{\nu}}} \right] .
\label{F(r)}
\end{eqnarray}
We ignore the rate of deposition energy reabsorved by the central black hole.
As was estimated in AF, the effect of reabsorption is not so large.

\section{RESULTS}

\subsection{Isothermal Disk}

We estimate EDR by neutrinos emitted from
the isothermal accretion disk.
For this case, neglecting the rotation of the disk
and the gravitation of the central black hole,
AF calculated EDR in a wide region apart from the
rotation axis as well as along it.
For the reference, EDR by $\nu_{\rm e}$
over $r=1.5$-$10 r_g$ and $\theta \leq \pi/4$ is
\begin{equation}
\frac{dE_0}{dt}=5.28 \times 10^{52} \left( \frac{k T_{\rm eff}(3 r_g)}
{10 {\rm MeV}} \right)^9
\left( \frac{r_g}{10 {\rm km}} \right)^3 \quad \mbox{ergs ${\rm s^{-1}}$},
\label{standard1}
\end{equation}
where $R_{\rm in}=3 r_g$ and $R_{\rm out}=10 r_g$.
In this case the energy fraction deposited in this region is
\begin{equation}
\frac{\dot{E_0}}{\cal L}=
2.03 \times 10^{-2} \left( \frac{k T_{\rm eff}(3 r_g)}{10 {\rm MeV}} \right)^5
\left( \frac{r_g}{10 {\rm km}} \right),
\label{frac}
\end{equation}
where ${\cal L}$ is the neutrino and antineutrino luminosity.
If $k T_{\rm eff}(3 r_g)$ is 10 MeV, about 2\% of the neutrino energy is
deposited in this region.
This temperature is almost the same as the typical temperature
in the simulation of neutron star mergers by Ruffert \& Janka (1999).
We have neglected the decrease of neutrino density due to
the annihilation in our formulation.
Thus the energy fraction gets larger than one if we adopt
a higher temperature than
$k T_{\rm eff}(3 r_g) \simeq 20$ MeV in equation (\ref{frac}).
This implies that neutrinos become optically thick for the pair
annihilation in this case.

In the following, we calculate EDR incorporating 
the disk rotation and the gravitation of the central black hole.
As a matter of convenience, AF assumed 
that the temperature of the accretion disk is relativistically isothermal,
namely, $T_0$ is constant.
In that case, we can estimate equation (\ref{depo}) by calculating the 
only two neutrino trajectories coming from $R_{\rm in}$ and $R_{\rm out}$.
$T_0$ goes out of the integral in equation (\ref{F(r)}).
However, it may be more natural to take
the effective temperature $T_{\rm eff}(R)$ as a base since the temperature
(and other thermodynamic quantities) is well defined in the comoving coordinates.
Thus in this subsection we assume the constant effective temperature,
$T_{\rm eff}(R)=T_{\rm eff}(3 r_g)$.
In this case $T_0$ becomes hotter as $R$ goes to outer 
region.
If we assume $T_{\rm eff}(3 r_g)$ is common, 
EDR for the constant $T_{\rm eff}$
becomes larger than that for the constant $T_0$.

Integration of equation (\ref{depo}) over the volume $dV=(r^2+a^2 \cos^2{\theta})
\sin{\theta} dr d\theta d\varphi$ gives the energy deposition per 
unit world time for a distant observer.
However, we estimate EDR within the infinitesimal
angle $d\theta$ along the rotation axis over $r=1.5$-$10 r_g$.
We do not consider EDR at $r < 1.5 r_g$
since in this region the baryon contamination occurs severely and the reabsorption
rate of the deposited energy to the black hole is large.
Then EDR is proportional to the dimensionless integral of
$G(r) \equiv F(r) (r^2+a^2)/r_g^2$ over $\hat{r} \equiv r/r_g$;
\begin{equation}
\frac{dE_0}{dt} \simeq 4.41 \times 10^{48}
\left( \frac{d \theta}{10^\circ} \right)^2
\left( \frac{k T_{\rm eff}(3 r_g)}{10 {\rm MeV}} \right)^9
\left( \frac{r_g}{10 {\rm km}} \right)^3 
\int_{1.5}^{10} G(\hat{r}) d \hat{r} \quad \mbox{ergs ${\rm s^{-1}}$},
\end{equation}
for $\nu_{\rm e}$.

We discuss the cases that the dimensionless quantity $a_* \equiv a/0.5 r_g$
takes the values $0$, $0.9$, and $0.99$.
We adopt the innermost stable orbit as the inner edge of the disk, $R_{\rm in}$.
As $a_*$ takes a larger value, $R_{\rm in}$ becomes smaller.
$R_{\rm in}$ are $3$, $1.16$, and $0.73 r_g$ for $a_*=0$, $0.9$, and $0.99$, respectively.
For $a_*=0.99$, $R_{\rm in}$ is inside of the ergosphere.
$R_{\rm out}$ is fixed at $10 r_g$ in every case.

We plot $G(r)$ in Figure 1 for all these cases.
This figure shows that EDR in the case of $a_*=0$,
the Schwarzschild black hole, is enhanced in comparison 
with the case ignoring the relativistic effects.
The disk temperature observed at infinity is higher at the outer region
unlike the model of constant $T_0$ discussed in AF.
Correspondingly EDR should be larger than the case of AF.
On the other hand, the Doppler effect due to the rotating disk reduces EDR.
Consequently the result is not so different from that of AF.

In Table 1 we list the integral of $G(r)$ over $r=1.5$-$10 r_g$.
When $R_{\rm in}=3 r_g$ and the relativistic effects are neglected,
electron neutrinos deposit in this region the energy,
\begin{equation}
\frac{dE_0}{dt} \simeq 2.23 \times 10^{51}
\left( \frac{d \theta}{10^\circ} \right)^2
\left( \frac{k T_{\rm eff}(3 r_g)}{10 {\rm MeV}} \right)^9
\left( \frac{r_g}{10 {\rm km}} \right)^3 \quad \mbox{ergs ${\rm s^{-1}}$}.
\label{standard2}
\end{equation}
The values listed in Table 1 are normalized by this EDR.
For $k T_{\rm eff}(3 r_g)=10$ MeV in this case,
the energy deposited in this region is 0.086 \% of the neutrino luminosity.
As is understood from Table 1 and Figure 1,
EDR becomes larger as $a_*$ takes a larger value.
The smaller $R_{\rm in}$ causes the larger neutrino luminosity.
Although the luminosity observed at infinity for $a_*=0.9$ is at most 10 \% larger
than the case for $a_*=0$,
EDR for $a_*=0.9$ becomes 1.3 times of the latter.
For $a_* \geq 0.99$ EDR changes little and practically remains constant.
This is because the disk surface does not change very much
and the neutrinos emitted from the innermost region suffer the large redshift
due to the disk rotation and gravitation for $a_* \geq 0.99$. 
As is seen from Figure 1, the peak of $G(r)$ approaches to the origin
as $a_*$ becomes larger.
This is not desirable in order to avoid the baryon contamination problem.
However, this negative property is overcome by the increase
of the deposited energy.

From the above discussion, we conclude that the relativistic effects
make increase EDR along the rotation axis by about factor two,
irrespective of $a_*$.
However, the order of EDR is unchanged.
Therefore, even if we calculate EDR
neglecting the relativistic effects,
we can estimate roughly the energy of gamma-ray bursts.

\subsection{Disk with Temperature Gradient}

EDR via neutrino pair annihilation is strongly
dependent on the disk temperature as $dE_0/dt \propto T_{\rm eff}^9$.
Thus the isothermal disk is too crude even for the zeroth-order approximation.
In this subsection, we discuss the pair annihilation of neutrinos emitted from 
the disk with temperature gradient.
Here, as before, we do not consider
the detailed structure and the formation process of the disk.
We have no definite theory on the temperature gradient of the accretion
disk as the source of gamma-ray bursts.
Here we adopt $T_{\rm eff}(R)=T_{\rm eff}(3 r_g) \cdot 3 r_g/R \propto R^{-1}$
as the simplest and acceptable model (see e.g. Kato, Fukue \& Mineshige 1998).
In this model the temperature profiles for different $a_*$ remain
common outside of $R=3 r_g$.
Actual accretion disks may not be opaque and not hot in the central region.
Here, however, we consider the idealized disk model mentioned above.
This disk model has the highest temperature at $R_{\rm in}$
where we need the most detailed survey of the gravitational effects.
The efficiency of the energy deposition is very high
because the innermost temperature is highest.

In this case it is not suitable unlike the isothermal case to discuss 
based on the disk of $R_{\rm in}=3 r_g$.
As $R_{\rm in}$ gets decreased,
the neutrino luminosity
becomes very large due to neutrinos
coming from the hotter region inside of $R=3 r_g$ and enhances EDR drastically.
For instance, in the case ignoring the gravitation and disk rotation,
EDR in the region $r=1.5$-$10 r_g$ and
$\theta \leq \pi/4$ is
\begin{eqnarray}
\frac{dE_0}{dt}&=&5.76 \times 10^{50} \left( \frac{k T_{\rm eff}(3 r_g)}
{10 {\rm MeV}} \right)^9
\left( \frac{r_g}{10 {\rm km}} \right)^3 \quad \mbox{ergs ${\rm s^{-1}}$ for
$R_{\rm in}=3 r_g$}, \\
&=&1.05 \times 10^{53} \left( \frac{k T_{\rm eff}(3 r_g)}
{10 {\rm MeV}} \right)^9
\left( \frac{r_g}{10 {\rm km}} \right)^3 \quad \mbox{ergs ${\rm s^{-1}}$ for
$R_{\rm in}=1.16 r_g$},
\end{eqnarray}
where we have adopted $R_{\rm out}=10 r_g$ as before.
These values have been calculated by use of the formulation in AF,
taking the temperature gradient into consideration.
For $R_{\rm in}=3 r_g$, in comparison with the isothermal case (see equation
(\ref{standard1})), the 
luminosity and EDR are reduced roughly by factor 10 and 100, respectively
for the same $T_{\rm eff}(3 r_g)$.
Therefore, EDR gets less efficient.
For $R_{\rm in}=1.16 r_g$, on the other hand, the luminosity is about 0.7
times that of the isothermal disk with $R_{\rm in}=3 r_g$. However, EDR
becomes about two times larger and the efficiency is improved 
compared with the isothermal disk for the same $T_{\rm eff}(3 r_g)$.
We have listed in Table 2 EDR along the rotation axis
over $r=1.5$-$10 r_g$ for various $R_{\rm in}$ in 
the case ignoring the relativistic effects.
$R_{\rm in}=1.45$ and $0.59$ correspond to the innermost stable orbit
for $a_*=0.8$ and $0.999$, respectively.
Table 2 shows that EDR increases rapidly due to the decrease
of $R_{\rm in}$.
The neutrino luminosity emitted from the disk with $R_{\rm in}=0.73 r_g$ becomes
only 19 times that of the case for $R_{\rm in}=3 r_g$, whereas EDR of the former case
increases by factor 1400 compared with the latter one.

In the following we estimate EDR taking into consideration
the gravitation of the central black hole and the disk rotation.
The relativistic effects are estimated comparing
with EDR for the same $R_{\rm in}$ without the relativistic effects.
We adopt $R_{\rm out}=10 r_g$ as before.
We plot $G(r)$ in Figure 2 for the Schwarzschild black hole.
In the case of $T_{\rm eff} \propto R^{-1}$,
EDR decreases rapidly with $r$ unlike the isothermal case.
EDR increases at small $r$ due to the gravitational lensing.
Since the neutrino energy decreases due to the gravitational redshift,
EDR is less efficient at $r$ larger than $r=6 r_g$ than that
without the relativistic effects.
As in the isothermal case,
the relativistic effects make the point of the maximum rate
approach to the origin.
As is shown in Table 3, the relativistic effects for $a_*=0$ enhance EDR by factor 1.7.
This is the same tendency as in the case of the isothermal disk.

Whereas, for $a_*=0.8$ it goes from Figure 3 that the relativistic 
effects enhance EDR only in the very neighborhood of the black 
hole and reduce it at distant places due to the redshift.
These relativistic effects are disadvantageous to avoid
the baryon contamination by the energy deposition at distant places
from the central object.
The peak of EDR lies in  $r < 1.5 r_g$.
Table 3 indicates that EDR for $a_*=0.8$ over $r=1.5$-$10 r_g$
is not changed very much compared with the case ignoring the relativistic effects.

As $a_*$ takes a larger value, EDR itself
increases.
For $a_* \geq 0.9$, however,
the energy reduction by the relativistic effects becomes larger as $a_*$ increases
(see Table 3).
For $a_*=0.9$, as is shown in Figure 4, $G(r)$ takes lower values than
those in the case ignoring the relativistic effects in almost all
regions and we cannot expect a large deposition at distant places from
the center.
In Figure 5, $G(r)$ for $a_*=0.99$ is much smaller than that ignoring
the relativistic effects in all region, especially at distant places.
These results are disadvantageous to avoid the baryon contamination problem
in gamma-ray bursts.
In these cases, EDR is largely owing to the neutrinos
emitted from the innermost hot region of the disk.
These neutrinos suffer the large redshift due to the rapid disk rotation and
gravitation and cannot effectively deposit energy at $r>R$.
Figure 5 clearly shows that EDR substantially increases when $R_{\rm in}$
decreases in the case neglecting the relativistic effects.
On the other hand,
if we plot G(r) for $a_*=0.999$ in Figure 5
taking into consideration the relativistic effects,
$G(r)$ overlaps almost completely with those for $a_*=0.99$ with the relativistic effects
(solid line).

The neutrino luminosity observed at infinity for, as an example, $a_*=0.99$ is 
reduced to 0.24 times that ignoring the relativistic effects.
Thus the redshift seriously affects the luminosity,
which results in Table 3.
However, the results of Table 3 cannot be explained
only by the change of the luminosity.
For when  $a_*$ goes from $0.99$ to $0.999$, the luminosity increases by
factor 1.2 but EDR remains almost constant.
Whereas if we neglect the relativistic effects, the luminosity
and EDR increase, respectively, by factor 1.5 and 2 for the same change
of $R_{\rm in}$.
The different pattern of the EDR growth between these two cases
indicates that the relativistic effects on the reaction rate of neutrino 
pair annihilation are also important.
Even if $a_*$ changes,
EDR at distant places is practically unchanged for $a_* > 0.8$.
Most of the energy growth due to the increase of $a_*$
occurs near the center.

As mentioned above, when $a_*$ goes beyond $0.99$, EDR itself remains
almost constant.
For $T_{\rm eff} \propto R^{-1}$, even if $a_*$ approaches to 1 beyond $0.99$,
the deposited energy does not increase because of the redshift due to the disk
rotation and gravitation unlike the case neglecting the gravitational effects.
The obtained EDR is fairly small compared with that in the case
where $R_{\rm in}$ approaches to $0.5 r_g$ and the relativistic effects
are neglected.
Namely, as $a_*$ approaches to one, the value of about 2.7 times equation
(\ref{standard2}) is the rough maximum value of EDR in this model.
The fraction of EDR against the neutrino luminosity increases
as $a_*$ becomes large.
For $a_*=0.99$, if we take account of the relativistic effects,
the neutrino luminosity observed at infinity reduces by factor 0.4 compared with the
case of equation (\ref{standard2}).
Therefore, the efficiency of EDR becomes 6.5 times that for the case of
equation (\ref{standard2}).
For the same $T_{\rm eff}(3 r_g)$,
Figure 1 and 5 indicate
that EDR at distant places above the disk
with $T_{\rm eff} \propto R^{-1}$ is much smaller than those above the isothermal disk,
although the integrated deposition rates are not so different.
Thus the isothermal disk is more advantageous than the disk
with $T_{\rm eff} \propto R^{-1}$ to avoid the baryon contamination problem.

We have calculated for the two extreme disk models,
$T_{\rm eff} \propto R^{0}$ and $R^{-1}$,
on behalf of various disk models.
The calculations for  $T_{\rm eff} \propto R^{-1}$ show
the following things:
If the deposited energy is mainly contributed by the neutrinos
coming from the central region,
the bending effect which increases EDR is dominant for large $R_{\rm in}$,
and the redshift effect which decrease EDR is dominant for small $R_{\rm in}$.
Namely, as $R_{\rm in}$ approaches the horizon,
the redshift effect overwhelms the bending effect.
If we assume a model with the more flat temperature gradient
than $T_{\rm eff} \propto R^{-1}$ ( the numerical work of Popham,
Woosley \& Fryer (1999) predicts such disks),
the contribution of the neutrinos coming from central region is relatively
diminished compared with the case of $T_{\rm eff} \propto R^{-1}$.
Therefore the increase of EDR due to the decrease of $R_{\rm in}$
is not so prominent as in the cases of Table 2 and 3,
and correspondingly the relativistic effects are not so manifest as shown in Table 3.
However, the qualitative result that the redshift effect becomes dominant
according to the decrease of $R_{\rm in}$ (or according to the growth of $a_*$)
is unchanged.
We have listed the relativistic effects for the disk with
$T_{\rm eff}(R)=T_{\rm eff}(3 r_g) \cdot \sqrt{3 r_g/R} \propto R^{-0.5}$
in Table 4.
This table strongly supports our consideration mentioned above.
Since the redshift effect becomes dominant as $a_*$ increases,
the enhancement factor of EDR due to the
relativistic effects has the maximum value in the Schwarzshild case.
The enhancement factor in Table 4 is 1.7,
which is the same as those in the former two disk models.
From these arguments it is conjectured that the enhancement factor due to the
relativistic effects probably does not exceed 2
in the disk models with simple temperature dependences.
Also in Table 4, for $a_* > 0.99$,
EDR itself remains almost constant as in the other two cases.

We are not able to investigate all possible models with
different temperature distributions. In order to study the
qualitative properties of relativistic effects we have
calculated only the simple models mentioned above. However, if
the opacity and the density in the central region are low, EDR
is not dominated by the neutrinos coming from that region. The
redshift effect is not so prominent as that of our cases. In
that case the role of $R_{\rm in}$ is replaced by the radius at
which $T_{\rm eff}$ has the highest value.  As this radius moves
outward, the  bending effect gradually becomes dominant and the
enhancement factor due to the relativistic effects may approach
1.7 of $a_*=0$ case.

\section{CONCLUSIONS}

\indent

In this article we have investigated
the relativistic effects on
EDR via neutrino pair annihilation
near the rotation axis of a Kerr black hole with the thin accretion disk.
The energy deposited over $r=1.5$-$10 r_g$ is our concern in this paper.
The bending of neutrino trajectories and the redshift due to
the disk rotation and gravitation have been taken into consideration.
The Kerr parameter, $a$, affects not only neutrinos' behavior but also
the inner radius of the accretion disk.
If the accretion disk is isothermal,
the relativistic effects make increase EDR by about factor two,
irrespective of $a$.

On the other hand, if the temperature of the accretion disk
behaves as $T_{\rm eff} \propto R^{-1}$,
the increase of $a$ enhances EDR drastically.
However, the most of the pair annihilation occur near the central black hole.
This is disadvantageous to avoid the baryon contamination problem
in gamma-ray bursts.
When $a_*$ goes beyond $0.99$, EDR remains almost constant
unlike the case without the relativistic effects                                             
and is much smaller than EDR in the case where $R_{\rm in}$ approaches to $0.5 r_g$
and the relativistic effects are neglected.
This is because the neutrinos,
emitted from the innermost hot spot of the disk,
suffer the large redshift.
This is the case also for the disk with the more flat temperature gradient
and for the isothermal disk.
Therefore, these results indicate that
EDR hardly increases, irrelevant to models of the disk temperature,
even if we consider the disk with small $R_{\rm in}$.

From the above arguments the qualitative properties
of the relativistic effects on neutrino
pair annihilation are summarized as follows.
When the deposition energy is mainly contributed by the neutrinos coming
from the central part, the redshift effect becomes dominant as this
active part approaches to the horizon and EDR is reduced compared with
that neglecting the relativistic effects.
On the other hand, as the part emitting the dominant neutrinos goes away
from the horizon, the bending effect gets dominant to make EDR increase
by factor 2 compared with that neglecting the relativistic effects.

These qualitative conclusions may be helpful for numerical
simulations of gamma-ray burst sources which neglect the
relativistic effects in propagation of neutrinos. In order to
discuss quantitatively the central engine of gamma-ray bursts, one
needs numerical simulations and several groups are acting in
this field (e.g. Popham, Woosley \& Fryer 1999; MacFadyen \&
Woosley 1999; Ruffert \& Janka 1999). The gravitational effects
on neutrino pair annihilation may quantitatively change the
results of such  simulations. As it comes from our calculations,
the estimated power of a gamma ray burst may be changed by a
factor less than two if the temperature gradient is relativelly
small, and the neutrinos emitted close to the inner adge of the
disk do not dominate. However, if $a_*$ is close to one and EDR
is strongly dominated by the neutrinos coming from the central
region, one can not give a reliable result without taking the
redshift effect into consideration.

\vspace{1.5cm}

We are greatly indebted to F. Takahara for useful discussion.
We are also grateful to S. Kobayashi and for helpful advice and comments.
This work was partly supported by a Research Fellowship of the Japan Society for
the Promotion of Science.

\newpage

\begin{center}
{\bf \LARGE References}
\end{center}

\medskip

\begin{description}

\item
Asano, K., \& Fukuyama, T. 2000, ApJ, 531, 949 (AF)
\item
Bardeen, J. M., Press, W. H., \& Teukolsky, S. A. 1972, ApJ, 178, 347
\item
Berezinsky, V. S., \& Prilutsky, O. F. 1987, A\&A, 175, 309
\item
Chandrasekhar, S. 1983, The Mathematical Theory of Black Holes
(New York: Oxford)
\item
Cooperstein, J., Van Den Horn, L. J., \& Baron, E. 1987, ApJ, 321, L129
\item
Covino, S., et al. 1999, A\&A, 348, L1
\item
Fruchter, A. S., et al. 1999, ApJ, 519, L13
\item
Goodman, J., Dar, A., \& Nussinov, S. 1987, ApJ, 314, L7
\item
Harrison, F. A., et al. 1999, ApJ, 523, L121
\item
Huang, Y. F., Dai, Z. G., \& Lu, T. 2000, A\&A, 355, L43
\item
Janka, H.-T., \& Ruffert, M. 1996, A\&A, 307, L33
\item
Jaroszy\'nski, M. 1993, ActaAstron, 43, 183
\item
Jaroszy\'nski, M. 1996, A\&A, 305, 839
\item
Kato, S., Fukue, J., \& Mineshige, S. 1998, Black-hole Accretion Disks
(Kyoto: Kyoto University)
\item
Landau, L. D., \& Lifshitz, E. M. 1979, Classical Theory of Fields
(London: Pergamon)
\item
MacFadyen, A., \& Woosley, S. E. 1999, ApJ, 524, 262
\item
M\'esz\'aros, P., \& Rees, M. J. 1992, MNRAS, 257, 29p
\item
M\'esz\'aros, P., \& Rees, M. J. 1993, ApJ, 405, 278]
\item
Paczy\'nski, B. 1990, ApJ, 363, 218
\item
Popham, R., Woosley, S. E., \& Fryer, C. 1999, ApJ, 518, 356
\item
Rees, M. J., \& M\'esz\'aros, P. 1992, MNRAS 258, 41p
\item
Ruffert, M., \& Janka, H.-T. 1998, A\&A, 338, 535
\item
Ruffert, M., \& Janka, H.-T. 1999, A\&A, 344, 573
\item
Ruffert, M., Janka, H.-T., Takahashi, K., \& Sch\"afer, G. 1997, A\&A, 319, 122
\item
Salmonson, J. D., \& Wilson, J. R. 1999, ApJ 517, 859
\item
Sari, R., Narayan, R., \& Piran, T. 1996 ApJ, 473, 204
\item
Sari, R., \& Piran, T. 1995, ApJ, 455, L143
\item
Shemi, A., \& Piran, T. 1990, ApJ, 365, L55
\item
Wijers, R. A. M. J., et al. 1999, ApJ, 523, L33
\item
Woosley, S. E. 1993, ApJ, 405, 273

\end{description}
\newpage

\begin{center}
{\bf \large Figure captions}
\end{center}

\figcaption{The plot of $G(r)$ (solid line)
against $r/r_g$ for the isothermal disk.
The dashed line corresponds to the case for $R_{\rm in}=3 r_g$
ignoring the relativistic effects.}

\figcaption{The plot of $G(r)$ (solid line)
against $r/r_g$ for the case of $a_*=0$ 
and $T_{\rm eff} \propto R^{-1}$. The dashed line corresponds to
the case for the same $R_{\rm in}$ ignoring the relativistic effects.
Contrary to Figure 1, the vertical axis is measured by the logarithmic scale.}

\figcaption{Same as Figure 2 but for $a_*=0.8$.}

\figcaption{Same as Figure 2 but for $a_*=0.9$.}

\figcaption{Same as Figure 2 but for $a_*=0.99$.
The dashed line and dotted line correspond to the cases for $R_{\rm in}=0.73$
($a_*=0.99$) and $0.59 r_g$($a_*=0.999$) ignoring
the relativistic effects, respectively.}

\newpage

\begin{center}
\begin{tabular}{cccc}
\hline \hline
$a_*$ & 0 & 0.9 & 0.99 \\ \hline
EDR & 1.7 & 2.2 & 2.3 \\ \hline
\end{tabular}
\end{center}
{\footnotesize Table~1. EDR within the infinitesimal
angle $d\theta$ along the rotation axis over $r=1.5$-$10 r_g$
for the isothermal disk.
The values are normalized to unity
when $R_{\rm in}=3 r_g$ and the relativistic effects are neglected.}

\begin{center}
\begin{tabular}{cccccc}
\hline \hline
$R_{\rm in}/r_g$ & 3 & 1.45 & 1.16 & 0.73 & 0.59 \\ \hline
EDR & 0.011 & 0.70 & 2.1 & 15 & 31 \\ 
${\cal L}$ & 0.09 & 0.41 & 0.65 & 1.7 & 2.5 \\ \hline
\end{tabular}
\end{center}
{\footnotesize Table~2. EDR along the rotation axis over
$r=1.5$-$10 r_g$ for the disk with 
$T_{\rm eff} \propto R^{-1}$ ignoring the relativistic effects.
The given values are normalized by the value of equation (\ref{standard2}).
The neutrino luminosity ${\cal L}$ are also listed.
Each ${\cal L}$ is normalized by the luminosity in the case of equation (\ref{standard2}).}

\begin{center}
\begin{tabular}{cccccc}
\hline \hline
$a_*$ & 0 & 0.8 & 0.9 & 0.99 & 0.999 \\
$R_{\rm in}/r_g$ & 3 & 1.45 & 1.16 & 0.73 & 0.59 \\ \hline
EDR & 1.7 (0.019) & 1.1 (0.79) & 0.73 (1.5)& 0.18 (2.6)& 0.087 (2.7)\\ 
${\cal L}$ & 0.62 (0.056)& 0.43 (0.18)& 0.36 (0.24)& 0.24 (0.40)& 0.19 (0.47) \\ \hline
\end{tabular}
\end{center}
{\footnotesize Table~3. EDR along the rotation axis over
$r=1.5$-$10 r_g$ for the disk with $T_{\rm eff} \propto R^{-1}$
including the relativistic effects. 
The neutrino luminosities ${\cal L}$ observed at infinity are also listed.
Each value is divided by that with the same $R_{\rm in}$
ignoring the relativistic effects.
The values normalized by the value
in the case of equation (\ref{standard2}) are also listed as parenthetic values.}

\begin{center}
\begin{tabular}{cccccc}
\hline \hline
$a_*$ & 0 & 0.8 & 0.9 & 0.99 & 0.999 \\
$R_{\rm in}/r_g$ & 3 & 1.45 & 1.16 & 0.73 & 0.59 \\ \hline
EDR & 1.7 (0.10) & 1.4 (0.40) & 1.2 (0.47)& 0.81 (0.54)& 0.68 (0.55)\\ 
${\cal L}$ & 0.69 (0.17)& 0.67 (0.23)& 0.66 (0.24)& 0.64 (0.33)& 0.63 (0.35) \\ \hline
\end{tabular}
\end{center}
{\footnotesize Table~4. Same as Table 3 but for $T_{\rm eff} \propto R^{-0.5}$.}

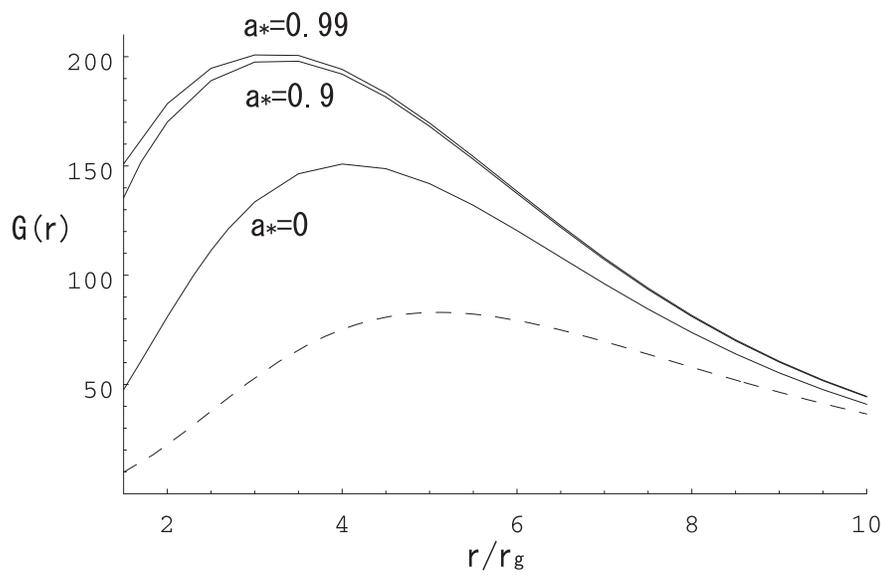
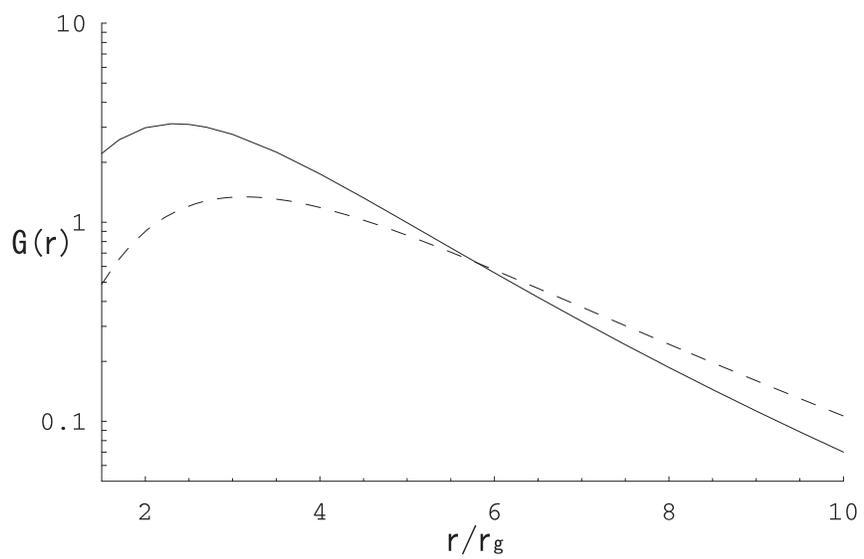
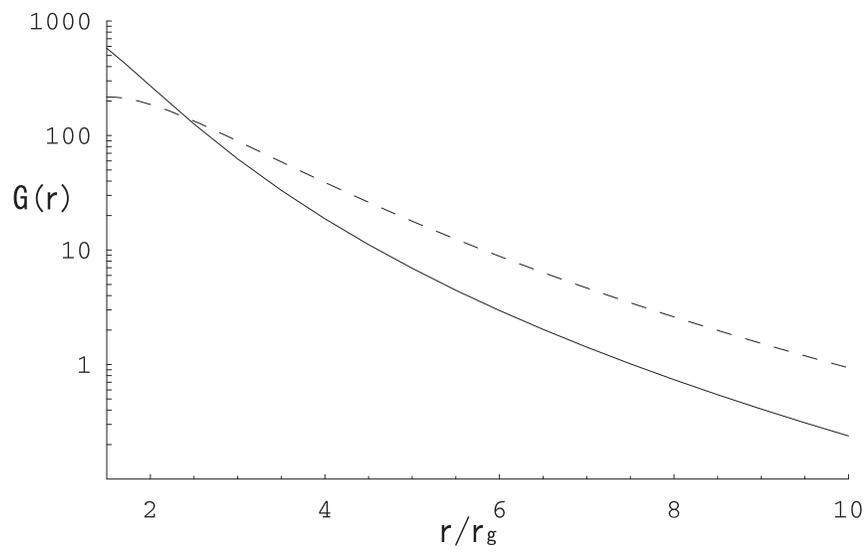
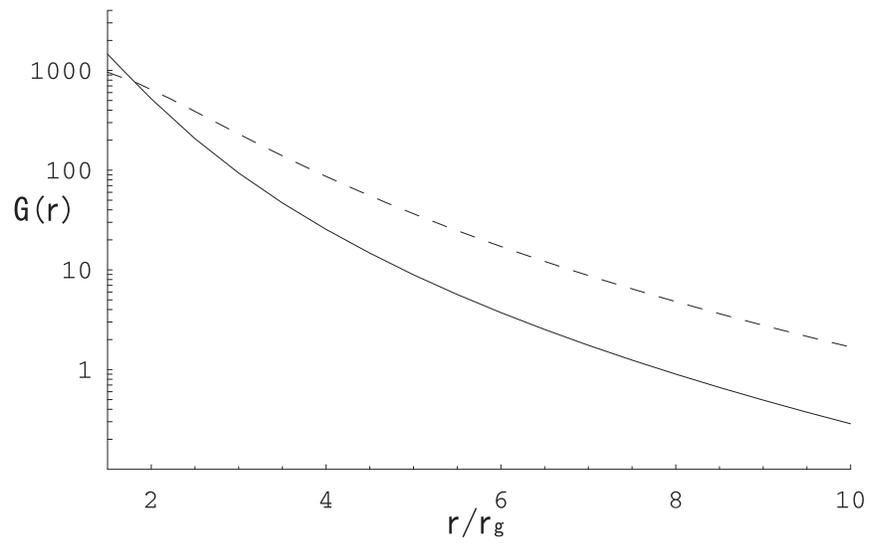
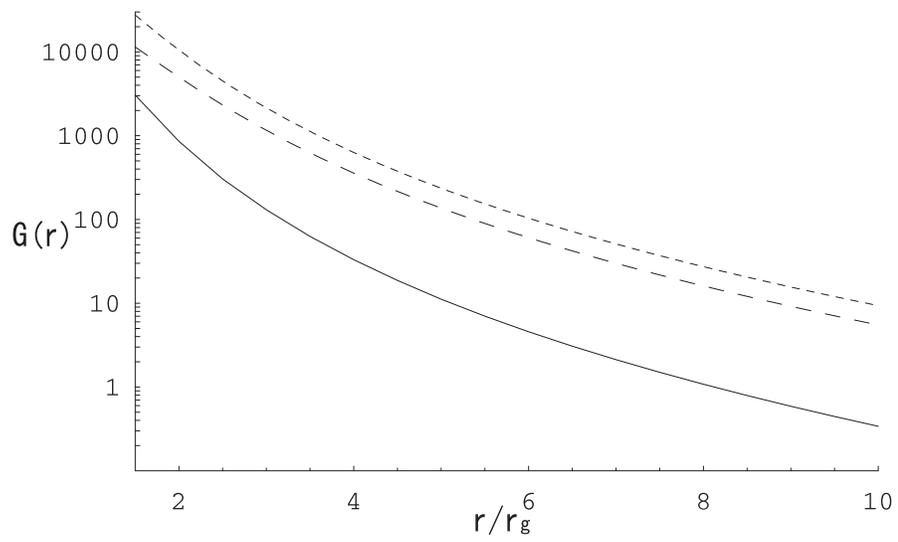
\end{document}